\documentclass[11pt]{article}

\usepackage{amsmath,amssymb,amsthm,amscd,latexsym}
\usepackage{hyperref}
\usepackage[toc,title,titletoc]{appendix}
\usepackage{cite}
\usepackage{graphicx,epsfig}
\RequirePackage{color}

\allowdisplaybreaks[1]

\def\virg#1{``#1''}

\def\eqi{\begin{equation}}
\def\eqf{\end{equation}}
\def\eqia{\begin{eqnarray}}
\def\eqfa{\end{eqnarray}}

\begin{document}

\title{Commentary  to \virg{LARES successfully launched in orbit: satellite and mission description}, by A. Paolozzi and I. Ciufolini}

\author{L. Iorio\\ Ministero dell'Istruzione, dell'Universit$\grave{\textrm{a}}$ e della Ricerca (M.I.U.R.)-Istruzione \\ Fellow of the Royal Astronomical Society (F.R.A.S.)\\ Viale Unit$\grave{\textrm{a}}$ di Italia 68, 70125, Bari (BA), Italy}

\maketitle

\begin{abstract}
We comment on some statements in a recent paper by Paolozzi and Ciufolini concerning certain remarks raised by us on the realistic accuracy obtainable in testing the general relativistic Lense-Thirring effect \textcolor{black}{in the gravitational field of the Earth} with the newly launched LARES satellite together with the LAGEOS and LAGEOS II spacecraft \textcolor{black}{in orbit} for a long time. The orbital configuration of LARES is  different from that of the originally proposed LAGEOS-3. Indeed, while the latter one should have been launched to the same altitude of LAGEOS (i.e. about $h_{\rm L}=5890$ km) in an orbital plane displaced by $180$ deg with respect to that of LAGEOS ($I_{\rm L}=110$ deg, $I_{\rm L3}=70$ deg), LARES currently moves at a much smaller altitude (about $h_{\rm LR}=1440$ km) and at a slightly different inclination ($I_{\rm LR} = 69.5$ deg). As independently pointed out in the literature by different authors, the overall accuracy of a LARES-LAGEOS-LAGEOS II Lense-Thirring test may be unfavorably \textcolor{black}{impacted} by the lower altitude of LARES with respect to the expected $\approx 1\%$ level  claimed by Ciufolini \textit{et al.} because of an enhanced sensitivity to the low-degree even zonal geopotential coefficients inducing orbital precessions competing with the relativistic ones. \textcolor{black}{Concerning the previous tests performed with the combined nodes of only LAGEOS and LAGEOS II, an independent analysis recently appeared in the literature indirectly confirms that the total uncertainty in them is likely far from being as little as $10\%$.}
\end{abstract}

\centerline{PACS: 04.80.-y; 04.80.Cc; 91.10.Sp; 91.10.Qm}
\centerline{Keyword 1: Experimental studies of gravity}
\centerline{Keyword 2: Experimental tests of gravitational theories}
\centerline{Keyword 3: Satellite orbits}
\centerline{Keyword 4: Harmonics of the gravity potential field}

\section{Introduction}
In a recent paper, Paolozzi and Ciufolini \cite{PC13} (PC13 hereafter) write: \virg{[\ldots] the concerns raised by one author, see,
e.g., \cite{IorioLares09b} are not based on solid grounds as clearly addressed and answered in a number of
papers, see, e.g., \cite{CiufoEPJ011,RiesEPL,CiufoNA012}  and references therein. The comments of that author \cite{IorioLares09b} in regard to
the errors due to the higher degree even zonal harmonics and to other conceivable error sources
in the LARES experiment are often even contradicting each other as pointed out, e.g., in \cite{CiufoNA012}.}
\section{Our comments}
Actually, in \cite{RiesEPL} the currently ongoing LARES mission along with the related issues is not treated at all: suffice it to say that even the very same word \virg{LARES}  is absent.

In regard to an alleged clear demonstration in  \cite{CiufoEPJ011} of the claimed unsoundness of the critical remarks raised in \cite{IorioLares09b}, nothing like that is actually present in \cite{CiufoEPJ011}, where neither \cite{IorioLares09b} nor any other paper by the present author is  cited.

PC13 did not cite other works \cite{Renzetti012,Renzetti013} on LARES which recently appeared in the literature criticizing some aspects of its proposed use to measure the Lense-Thirring drag and other fundamental physics effects.
In \cite{Renzetti012}, there is an analysis on the impact of the even zonals of higher degree on the LARES mission which basically supports the concerns raised in \cite{IorioLares09b} and in other works by the present author and co-workers about the accuracy of a Lense-Thirring test. Moreover,  Renzetti in \cite{Renzetti012} discusses various replies by Ciufolini \textit{et al.} to Iorio \textit{et al.} published in the literature by writing: \virg{[\ldots] the LARES
team did not support their rebuttals either by disclosing the
errors attributed to Iorio or by explicitly describing the procedure
adopted to obtain their own opposite outcome. In particular,
Ciufolini and co-workers [\ldots]
disclosed neither the approximations used nor the computational
approach followed; instead, it would be important to
know how many zonals were included in their calculations,
if they used the standard Kaula approach, as Ciufolini did in
the past, or if they adopted a different strategy, if their computations
are analytical or numerical, etc. [\ldots] In \cite{CiufoNA012} it seems
that they [i.e. Ciufolini \textit{et al.}] made new analyses contradicting the claims by Iorio
\textit{et al.} [\ldots], but actually it is not so as, instead, they repeated
statements previously published elsewhere. [\ldots] it is stated
that they [i.e. Ciufolini \textit{et al.}] used the model EIGEN-GRACE02S [\ldots] up to
$\ell = 10$; Ciufolini et al. [\ldots] claim that the inclusion of
higher degree even zonal harmonics would only change their
results slightly. This fact suggests that, contrary to what was
claimed, Ciufolini and co-workers actually did not deal with
the critical remarks raised by Iorio and co-workers.}.   We agree with the analysis presented in \cite{Renzetti012} by remarking that, once again, also in PC13 no detailed and explicit proofs of the claimed  lacking of \virg{solid grounds} in our analysis can be found.
In \cite{Renzetti013}, other aspects of using LARES for testing general relativity and fundamental physics are criticized as well.

\textcolor{black}{Moreover}, with regard to \virg{contradicting} statements, we can not help but notice how Ries \textit{et al.} \cite{RiesEPL} implicitly offers a negative assessment \textcolor{black}{of} the value of the currently ongoing LARES mission. \textcolor{black}{According to its proponents, it} should improve the tests\footnote{For a recent review, see \cite{RenzettiCEJP}.} \textcolor{black}{so far} performed with the LAGEOS and LAGEOS II satellites by one order of magnitude\textcolor{black}{, i.e. from $10\%$ to $1\%$}. \textcolor{black}{Now}, Ries \textit{et al.} \cite{RiesEPL}, just a few months before the launch of LARES, conclude by writing: \virg{The only way to make the experiment [i.e. the one made so far with LAGEOS and LAGEOS II] better would be to have launched LAGEOS-2 into the original LAGEOS-3 supplementary orbit.}

\textcolor{black}{Finally, we briefly comment on an independent result on the LAGEOS-LAGEOS II tests recently published in the literature \cite{2013MNRAS.432.2591P}. Fig. 1 of \cite{2013MNRAS.432.2591P} displays the time series $\delta\mu$ of the combined residuals of the nodes of LAGEOS and LAGEOS II, normalized to the expected Lense-Thirring signal, over a multidecadal time span starting in 1993. Let us recall that the  node residuals should entirely contain the unmodelled Lense-Thirring signature, along with the mismodelled effect due to the uncancelled even zonals of degree $\ell = 4,6,\ldots$ according to some  Earth's global gravity field model, left unspecified\footnote{\textcolor{black}{It may be EIGEN-GRACE02S \cite{eigengrace02s}.}} in Fig.  1 of \cite{2013MNRAS.432.2591P}. The largest biasing signal due to the first even zonal of degree $\ell = 2$ is cancelled, to an extent which is the main subject of \cite{2013MNRAS.432.2591P}, by construction. The average value of the time series in Fig.  1 of \cite{2013MNRAS.432.2591P}, depicted as a red horizontal straight line, is $\overline{\delta\mu} = 1.056$, while the expected value in general relativity is $\delta\mu_{\rm LT} = 1$. Interestingly, the scatter of the time series in Fig.  1 of \cite{2013MNRAS.432.2591P} yields an uncertainty which is certainly not as small as $0.1$, as it should be if the claims by Ciufolini \textit{et al.} concerning a $10\%$ total error were correct. From a direct visual inspection of Fig.  1 of \cite{2013MNRAS.432.2591P}, it seems that most of the values of $\delta\mu$ lie between 0 and 2, with some points reaching $-2$ and $4$. Unfortunately, the author of \cite{2013MNRAS.432.2591P} did not offer a quantitative evaluation of the uncertainty of $\delta\mu$. In any case, it should not be too dissimilar from the evaluations proposed by us in the literature, if not even larger.
}



\bibliography{replybib}{}

\end{document}